\documentclass{iopart}
\usepackage{psfig}
\begin{document}

\def\trace{{\rm Tr}\ \!}

\title{Numerical investigation of the aging
of the Fully-Frustrated XY model}

\author{J-C. Walter and C. Chatelain}
\address{
Groupe de Physique Statistique,\\
Institut Jean Lamour, UMR 7198,\\
Nancy-Universit\'e, CNRS,
BP~70239, Boulevard des aiguillettes,\\
F-54506 Vand{\oe}uvre l\`es Nancy Cedex, France}
\ead{chatelai@lpm.u-nancy.fr}

\begin{abstract}
We study the out-of-equilibrium dynamics of the fully-frustrated XY model.
At equilibrium, this model undergoes two phase transitions at two very close
temperatures: a Kosterlitz-Thouless topological transition and a second-order
phase transition between a paramagnetic phase and a low-temperature phase
where the chiralities of the lattice plaquettes are anti-ferromagnetically
ordered. We compute by Monte Carlo simulations two-time spin-spin
and chirality-chirality autocorrelation and response functions.
From the dynamics of the spin waves in the low temperature phase, we
extract the temperature-dependent exponent $\eta$. We provide evidences for
logarithmic corrections above the Kosterlitz-Thouless temperature and interpret
them as a manifestation of free topological defects. Our estimates of the autocorrelation
exponent and the fluctuation-dissipation ratio differ from the XY values,
while $\eta(T_{\rm KT})$ lies at the boundary of the error bar.
Indications for logarithmic corrections at the second-order critical
temperature are presented. However, the coupling between angles and
chiralities is still strong and explains why autocorrelation exponent and
fluctuation-dissipation ratio are far from the Ising values and seems stable.
\end{abstract}


\def\build#1_#2^#3{\mathrel{
\mathop{\kern 0pt#1}\limits_{#2}^{#3}}}

\section*{Introduction}
The 2D fully-frustrated XY model (FFXY) has attracted the attention of
several communities. Indeed, antiferromagnets on a triangular lattice
with a planar anisotropy enter in the class of FFXY models. The FFXY
also describes the superconducting-to-normal phase transition in
Josephson-junction arrays in a perpendicular magnetic field for half
a quantum of flux through each plaquette~\cite{Teitel82a}.
In this work, we consider the following Hamiltonian
	\begin{equation}
	H=-\sum_{(i,j)} (-1)^{x_ix_j} \cos(\theta_i-\theta_j)
	\end{equation}
where the sum extends over nearest neigbouring sites on the square lattice
and $(x_i,y_i)$ are the lattice coordinates of the $i$-th site.
This Hamiltonian corresponds to the so-called {\sl Pileup-Domino}
ordering of the ferromagnetic and anti-ferromagnetic couplings. Frustration
arises from the existence of three ferromagnetic and one anti-ferromagnetic
couplings in each plaquette.

\begin{center}
\begin{figure}
        \centerline{\psfig{figure=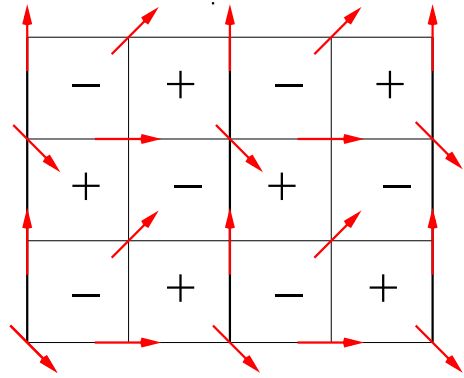,height=5cm}}
        \caption{Example of ground state of the FFXY. The thin lines are
	ferromagnetic bonds while the bold ones are antiferromagnetic.
	The plus and minus signs at the centers of the plaquettes are
	the chiralities.}
        \label{fig0}
\end{figure}
\end{center}

The phase diagram and critical properties of the FFXY has been subject of
debate for the last three decades. It is now established that the system
undergoes two phase transitions: a Kosterlitz-Thouless topological
transition~\cite{Kosterlitz73} associated to the breaking of the invariance
under the global rotation of all spins ($U(1)$ symmetry group) and a
second-order phase transition between a paramagnetic phase and a
low-temperature phase where chiralities of the spins around each plaquette
are anti-ferromagnetically ordered. This second transition is associated to
the breaking of a $Z_2$ symmetry (reversal of all chiralities) like in the Ising model.
The existence of two phase transitions has been supported by many Monte Carlo
simulations since the pioneering work of Teitel and Jayaprakash~\cite{Teitel82}.
The same conclusion has been drawn for the $J_1-J_2$ XY model~\cite{Simon}.
Theoretical arguments also predict a disentanglement of $U(1)$ and $Z_2$
degrees of freedom. Using a Hubbard-Stratonovich transformation, Choi and Doniach
showed the equivalence of the FFXY with two coupled O(2)-models~\cite{Choi85}.
The coupling is relevant under Migdal-Kadanoff renormalization
and in the strong coupling limit, the model reduces to a Ising-XY model
(the Ising spin being related to the relative orientation of the two $O(2)$-models).
Such a model have also been obtained by a mapping of the FFXY onto a
19-vertex model~\cite{Knopps94}. A modified version of the FFXY,
introduced by Villain, has allowed for the clarification of
the nature of the topological defects causing the Kosterlitz-Thouless transition.
By integrating out the spin-waves to leave only the contribution
of the topological defects, this model reduces to a gas of charged particles
($q=\pm 1/2$) in a uniform medium of charge $1/2$ interacting via a
logarithmic Coulomb potential~\cite{Villain77,Fradkin78}.
\\

A point of debate in the literature is whether the two phase transitions
occur at the same temperature or at two very close ones.
The breaking of the $Z_2$-symmetry is associated to a second-order phase
transition~\cite{Nussinov} caused by the proliferation of domain walls between
domains of different antiferromagnetic orderings of the chirality. As predicted by
Korshunov~\cite{Korshunov01} and then observed numerically~\cite{Olsson03},
at the kinks of these domain walls lies a $q=\pm 1/2$ topological defect
and the formation of dipoles of two opposite charges $q=+1/2$ and $q=-1/2$
stabilizes the domain wall. As a consequence, the proliferation of
these walls occur at higher temperatures. Recent Monte Carlo simulations
confirm that the two phase transitions occur at two distinct temperatures
(see \cite{Hasenbusch05} and references therein).
\\

Another point of debate concerns the universality class of the second-order
phase transition. Since the broken symmetry is $Z(2)$, this universality class
may be the same as the 2D Ising model. This idea is supported by the fact
that the coupling in the Ising-XY model has been shown to be irrelevant at the
critical point~\cite{Knopps94}. However, most of the Monte Carlo simulations for
the FFXY or the Ising-XY model conclude that $\nu\simeq 0.8$ while $\nu=1$ for
the Ising model (for a review of Monte Carlo estimations of the critical exponents,
see Table 1 of ref. \cite{Hasenbusch05}). The most recent Monte Carlo
simulations~\cite{Olsson95,Noh02,Hasenbusch05b, Hasenbusch05} now show evidences in
favor of the Ising universality class. The disagreement with older simulations
may be due to the fact that the exponents tend to the expected Ising values
only for large lattice sizes~\cite{Hasenbusch05b,Hasenbusch05}.
\\

In this paper, we are interested in the out-of-equilibrium behavior of the
FFXY. The short-time regime has been extensively studied~\cite{Luo97,Luo98b,Luo98a,Luo98}
in the aim to determine the static critical exponents as well as the dynamical exponents
$z$ and $\theta$. The growth of the correlation length, both for the angles and the
chiralities, has also been studied by numerical integration of the Langevin
equation~\cite{Kim95,Jeon03}. Temperature-dependent dynamical exponent were obtained for
both angles and chiralities, in contradistinction to short-time dynamics simulations and
relaxation of one-time quantities~\cite{Zheng03} for which $z$ remains close to $z\simeq 2$.
In the following, we present numerical evidences that both angles and chiralities
display simple aging like homogeneous ferromagnets. This implies in particular
that the dynamical exponent for the angles is $z=2$ for any temperature,
like in the XY model. As usual in the context of aging, we study the two-time
correlation and response functions from which we define a
fluctuation-dissipation ratio measuring the degree of violation of the
fluctuation-dissipation theorem. We consider two different protocols:
in the first section, the system is prepared in one of its ground state
and then let evolved below or at the Kosterlitz-Thouless temperature. The scaling
of the two-time functions are compared to the XY model. In the second section,
the system is prepared in the paramagnetic phase and then quenched at the
Kosterlitz-Thouless temperature. In the third section, the aging of the chirality
is studied when the system is prepared initially in the paramagnetic phase and
then quenched at the $Z_2$ critical temperature. 

\section{Spin-wave excitations at low temperatures}
\subsection{Spin-spin correlation function in the low-temperature phase}
The system is initially prepared in one of its ground states (figure \ref{fig0})
and then let evolved in the critical phase, i.e. at $T<T_{KT}=0.4461(2)$~\cite{Hasenbusch05},
with the Glauber dynamics~\cite{Glauber63}. If the coupling with the chirality
can be neglected out-of-equilibrium,
the thermal excitations associated to the angles are spin waves that should develop
in the system in the very same way than in the XY model but with an effective
coupling constant $J/\sqrt 2$~\cite{Choi85}. Previous Monte Carlo estimates of the
dynamical exponents found to be close to $z=2$~\cite{Luo98} (denoted $z_1$ in
this paper) is compatible with this picture. The two-time spin-spin correlation
function of the FFXY are thus expected to decay as in the XY model~\cite{Bray95}
	\begin{equation}
	C(t,s)=\sum_i \langle \vec\sigma_i(t).\vec\sigma_i(s)\rangle
	\sim (t-s)^{-\eta/2}\left({t+s\over\sqrt{ts}}\right)^{\eta/2}
	\label{eq1}
	\end{equation}
in the spin-wave approximation. The first term is due to short-time
reversible processes for which invariance under time translation is
manifest in the dependence on $t-s$ only (quasi-equilibrium regime). 
The second term is the contribution of long-time irreversible
processes causing aging. It depends only on $t/s$.
We have computed numerically these correlations for different final
temperatures $T=0.2, 0.4, 0.6$ and $0.8T_{KT}$. We have considered the waiting times
$s=200, 400, 600, 800, 1200, 1600, 2400, 3200, 4800, 5600,$ and $6400$.
The time $t$ runs from $s$ to $60.000$. The lattice size is
$256\times 256$ and the data have been averaged over 12,000 histories.
\\

\begin{center}
\begin{figure}
        \centerline{\psfig{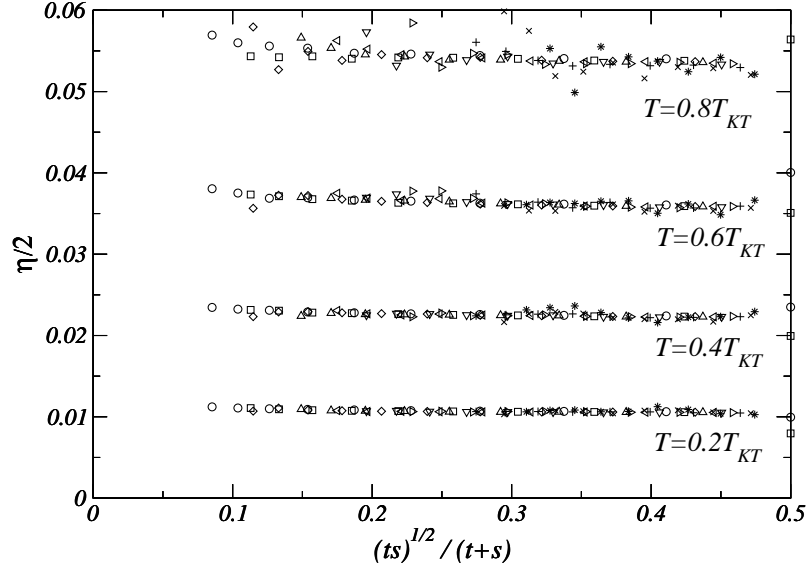}}
        \caption{Effective exponent $\eta/2$ obtained by interpolation of the
	spin-spin correlation function in the quasi-equilibrium regime.
	The different curves correspond to temperatures $0.2T_{KT}$, $0.4T_{KT}$, $0.6T_{KT}$
	and $0.8T_{KT}$ (from bottom to top). The different symbols correspond to
	different interpolation ranges from $s\ge 200$ (circle) to $s\ge 5600$ (star).}
        \label{fig1}
\end{figure}
\end{center}

\begin{center}
\begin{figure}
        \centerline{\psfig{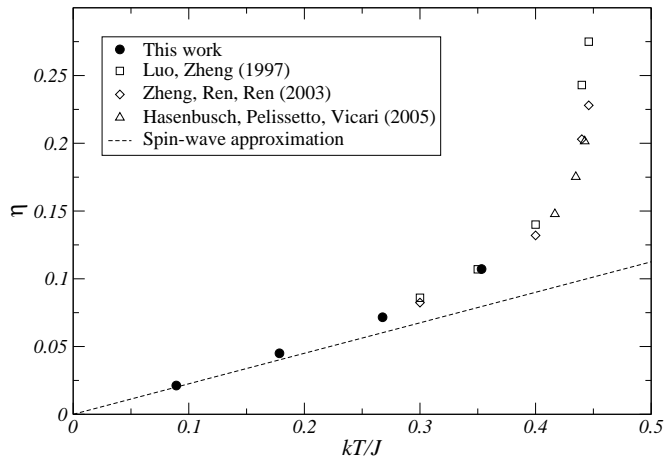}}
        \caption{Comparison of the exponent $\eta$ with other estimates found
	in the literature and the prediction of the spin-wave approximation.}
        \label{fig2}
\end{figure}
\end{center}

In a first step, we have interpolated the quasi-equilibrium regime
$(t-s)^{-\eta/2}$ to extract the static exponent $\eta$. To eliminate the
aging part, we analyzed the data along the curves
	\begin{equation}
	{t+s\over\sqrt{ts}}=\kappa={\rm Cste}
	\label{eq2}
	\end{equation}
On these curves, the correlation function $C(t,s)$ decay as $(t-s)^{-\eta/2}$
so that an estimation of $\eta$ is easily obtained by a linear fit of
$\ln C(t,s)$ versus $\ln (t-s)$. Practically, this was accomplished by first
doing an interpolation of the correlation function in the plane $(t,s)$
between the simulation points. On figure~\ref{fig1}, our estimates of $\eta/2$
are plotted versus $1/\kappa$. Since we are considering the quasi-equilibrium
regime, $t\sim s$, the value of $\eta$ has to be measured in the regime of
small $\kappa$ and thus on the right of the figure. Deviations, visible on the
left part of the figure, are due to corrections to the dominant behavior
of the correlation function. To test for the possible
existence of transient regime at small $s$, we have also tried to remove the
smallest waiting times: the different symbols on the figure correspond
to effective exponents taking into account all waiting times (circles), only
waiting times $s\ge 400$ (squares), $s\ge 600$ (diamonds) up to $s\ge 5600$
(stars). Even though the statistical errors increase when fewer waiting times
$s$ are taken into account, all curves are nicely compatible. Our final
estimates of $\eta$ are given in Table \ref{Table1}. They are in good agreement
with other estimates found in the literature (see figure \ref{fig2}), including
static Monte Carlo simulations, and close at low temperature to the linear
behavior $\eta={k_BT\sqrt 2\over 2\pi J}$ predicted in the spin-wave approximation.

\begin{table}[!ht]
\begin{center}
\begin{tabular}{@{}*{3}{l}}
$T/T_c$ & $\eta/2$ (MC) & $\eta/2$ (SW)\\
\hline
$0.2$ & $0.0106(6)$ & $0.0100$ \\
$0.4$ & $0.0225(9)$ & $0.0201$ \\
$0.6$ & $0.0358(8)$ & $0.0301$ \\
$0.8$ & $0.0536(20)$ & $0.0402$ \\
\hline
\end{tabular}
\end{center}
\caption{Comparison of the Monte Carlo estimates (MC) of the exponent $\eta/2$
obtained in the quasi-equilibrium regime with the predicted values in the
spin-wave approximation (SW).}
\label{Table1}
\end{table}

\begin{center}
\begin{figure}
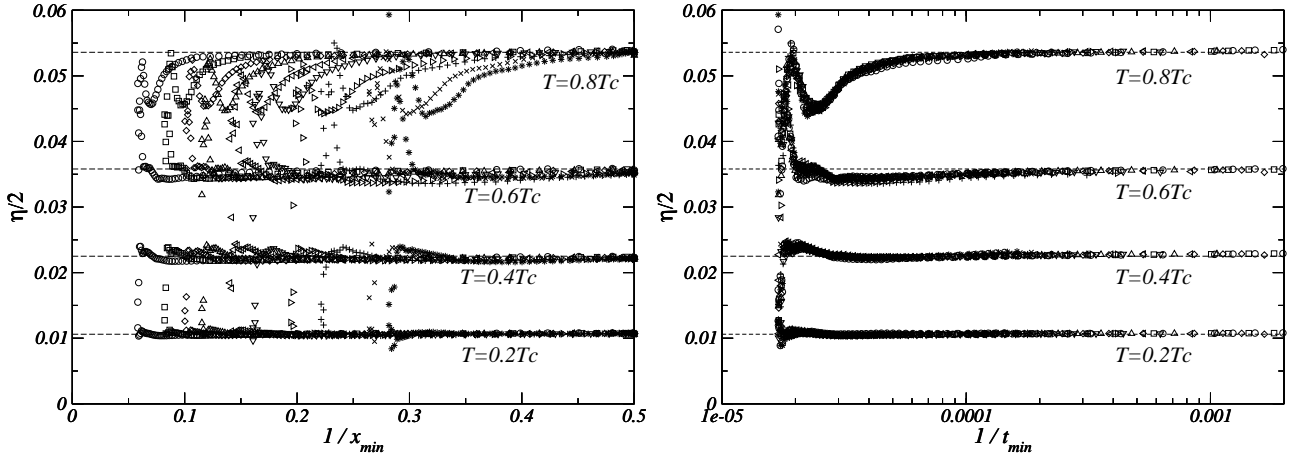

        \centerline{\psfig{figure=Corr-U1-Aging.eps,height=6cm}
	\psfig{figure=Corr-U1-Aging-v2.eps,height=6cm}}
        \caption{Effective exponent $\eta/2$ obtained by interpolation of the
	spin-spin correlation function in the aging regime. The different
	curves correspond to temperatures $0.2T_c$, $0.4T_c$, $0.6T_c$ and
	$0.8T_c$ (from bottom to top) and the different symbols to
	different waiting times $s$. The exponents $\eta/2$ obtained in the
	quasi-equilibrium regime are plotted as a dashed line. On the right,
	the same data are plotted with respect to $1/t_{\rm min}$.}
        \label{fig3}
\end{figure}
\end{center}

The exponent $\eta$ can be estimated from the aging part of the correlation
function $C(t,s)$ too. Since it is necessary to take into account correlation functions
at well separated times $s$ and $t$, the correlation function is very small and thus noisy.
A precise determination of $\eta$ is more difficult than in the quasi-equilibrium regime.
As a consequence, we will restrict ourselves to check that the previously estimated
exponents $\eta$ are compatible with our data in the aging regime. To that purpose, we
made a log-log interpolation of $(t-s)^{\eta/2}C(t,s)$ versus ${t+s\over\sqrt{ts}}$.
To take into account possible corrections to the asymptotic behavior, we varied
the interpolation window. Effective exponents are plotted on figure~\ref{fig3} with
respect to the smallest value of $x={t+s\over\sqrt{ts}}$ entering into the fit. The
rightmost points are obtained by an interpolation over all times $t\in [s+1;60,000]$.
The leftmost points come from an interpolation over only the three last times
$t\in [59,998;60,000]$. The different symbols correspond to different waiting
times $s$. The effective exponents display a plateau for the values obtained
in the quasi-equilibrium regime and plotted as a dashed line on figure~\ref{fig3}.
However, as temperature increases, the width of this plateau shrinks. It could
tempting to interpret the deviations from the plateau as the effect of
corrections to the aging behavior $x^{\eta/2}$. However, as can be seen on
the right figure~\ref{fig3}, where the effective exponents are now plotted
with respect to the smallest time $t$ entering into the fit, these deviations
depend only on $t$ and occur for large values of $t$. They are thus due to
an under-sampling of the correlation at large times $t$ that induces large
systematic fluctuations of the effective exponents when only the last few
points enter into the interpolation. They are thus numerical and not physical.

\begin{center}
\begin{figure}
        \centerline{\psfig{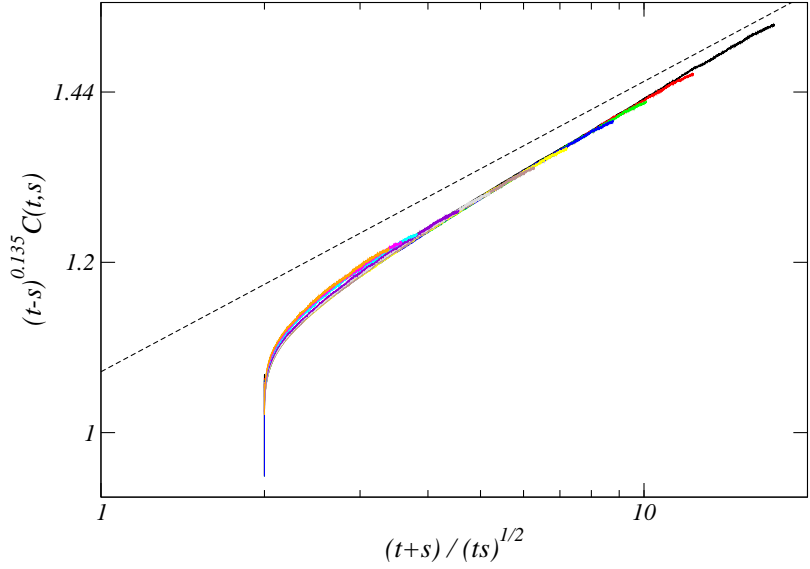}}
        \caption{Scaling function $(t-s)^{\eta/2}C(t,s)$
	of the spin-spin correlation function versus the scaling variable
	$(t+s)/\sqrt{ts}$ for the $FFXY$ model prepared in the ground state
	and quenched at its critical temperature. The different curves correspond
	to different waiting times ($s=200, 400, 600, 800, 1200, 1600, 2400, 3200,
	4800, 5600,$ and $6400$). The dashed curve is the expected asymptotic
	behavior $\big[(t+s)/\sqrt{ts}\big]^{\eta/2}$.}
        \label{fig4}
\end{figure}
\end{center}

In the case of a quench at the Kosterlitz-Thouless temperature $T_{KT}$, the estimation of the
exponent $\eta$ using the interpolation of $C(t,s)\sim (t-s)^{-\eta/2}$ in the
quasi-equilibrium regime becomes much less accurate. We obtain an estimate
$\eta\simeq 0.27(2)$. However, a nice collapse of the scaling function
$(t-s)^{\eta/2}C(t,s)$ versus the scaling variable $(t+s)/\sqrt{ts}$ is observed
as shown on figure \ref{fig4} and gives evidences of the validity of the expression
(\ref{eq1}) up to the Kosterlitz-Thouless transition temperature. The effective exponents
do not present any plateau but only seem to tend towards the expected value.

\subsection{Response function in the low-temperature phase}
The linear response of the observable $m$, say the magnetization, at time $t$
to an external field $h$, say the magnetic field, coupled to the system
at the instant $s$ is defined as
	\begin{equation}
	R(t,s)=\left({\delta\langle m(t)\rangle\over \delta h(s)}\right)_{h\rightarrow 0}
	\label{eq3}
	\end{equation}
At equilibrium, this quantity is related to the correlation function
$C(t,s)=\langle m(t)m(s)\rangle-\langle m(t)\rangle\langle m(s)\rangle$
by the fluctuation-dissipation theorem (FDT)
	\begin{equation}
	k_BT R(t,s)={\partial\over\partial s}C(t,s)
	\label{eq4}
	\end{equation}
The violation of this theorem in aging systems has attracted much attention
these last ten years. In the context of mean-field spin-glasses, the following
extended relation was introduced~\cite{Cugliandolo93}
	\begin{equation}
	k_BT R(t,s)=X(t,s){\partial\over\partial s}C(t,s)
	\label{eq5}
	\end{equation}
where $X(t,s)$ differs from the value $X=1$ out of equilibrium. In the aging
regime, $X(t,s)$ is believed to tend towards a universal quantity, denoted $X_\infty$.
\\

In the case of the FFXY model, the local order parameter is the two-dimensional
vector $\vec\sigma_i(t)$. As a consequence, correlation and response functions
are tensors:
	\begin{equation}
	C_{\mu\nu}(t,s)=\langle (\vec\sigma_i(t).\vec e_\mu)
	(\vec\sigma_i(s).\vec e_\nu)\rangle-\langle (\vec\sigma_i(t).\vec e_\mu)
	\rangle\langle(\vec\sigma_i(s).\vec e_\nu)\rangle
	\label{eq7}
	\end{equation}
	\begin{equation}
	R_{\mu\nu}(t,s)={\delta\langle\vec\sigma_i(t)\rangle
	\over\delta h^\nu(s)}.\vec e_\mu
	\label{eq8}
	\end{equation}
where the magnetic field is decomposed as $\vec h=h^\nu\vec e_\nu$. In the
following, we will consider only the traces $\trace C(t,s)$ and $\trace R(t,s)$
corresponding to the usual expression of the correlation function:
	\begin{eqnarray}
	\trace C(t,s)&=&\langle\cos\theta_i(t)\cos \theta_i(s)\rangle
	+\langle\sin\theta_i(t)\sin \theta_i(s)\rangle		\nonumber\\
	&=&\langle\cos(\theta_i(t)-\theta_i(s))\rangle
	\label{eq9}
	\end{eqnarray}
where $\vec s_i=\pmatrix{\cos\theta_i & \sin\theta_i}$.
This choice was already made in the study of the fluctuation-dissipation
ratio in the XY model~\cite{Abriet04}.
\\

Since the FFXY behaves at low temperature as an XY model with a stiffness
$J/\sqrt 2$, the response function can be calculated in the spin-wave
approximation. This leads to~\cite{Berthier01}
	\begin{equation}
	R(t,s)\sim (t-s)^{-1-\eta/2}\left({t+s\over
	\sqrt{ts}}\right)^{\eta/2}
	\label{eq6}
	\end{equation}
The response function can be computed directly in Monte Carlo
simulations~\cite{Chatelain03}. In principle, it could be used
to estimate $\eta$ from (\ref{eq6}). In practise, the response function
is noisier than the correlation function and thus does not allow for
a precise determination of $\eta$. However, with a sufficiently large
number of histories, the fluctuation-dissipation ratio $X(t,s)$
can be computed with a reasonable accuracy.
Using the scaling behavior for the correlation function (\ref{eq1})
and the response function (\ref{eq6}) in the spin-wave approximation, one can
calculate the scaling behavior of the fluctuation-dissipation ratio:
	$$X(t,s)={k_BTR(t,s)\over\partial_sC(t,s)}
	\sim {s(t+s)\over t^2-4ts-s^2}$$
that displays a divergence for $t/s=2+\sqrt 5$. Such a divergence was already
reported~\cite{Abriet} for the XY model. We also observe a divergence of $X(t,s)$
for the FFXY but its location tends to increase with the waiting time
$s$. Unfortunately, the computation of the fluctuation-dissipation ratio
is very CPU time-demanding and we had to restrict ourselves to a lattice
size $192\times 192$ and a final time $t_f=4000$. The data have been averaged
over $15,000$ histories at the two lowest temperatures and $45,000$ for the others.
The waiting times are $s=60, 80, 100, 120, 140$, and $160$.
As shown on figure \ref{fig6}, the location of the divergence $t_0/s_0$
displays a nice linear behavior with $1/s$ with a value at
$1/s\rightarrow 0$ compatible with the predicted value $2+\sqrt 5\simeq
4.236$. This provides an indirect confirmation of
the scaling behavior (\ref{eq6}).

\begin{center}
\begin{figure}
        \centerline{\psfig{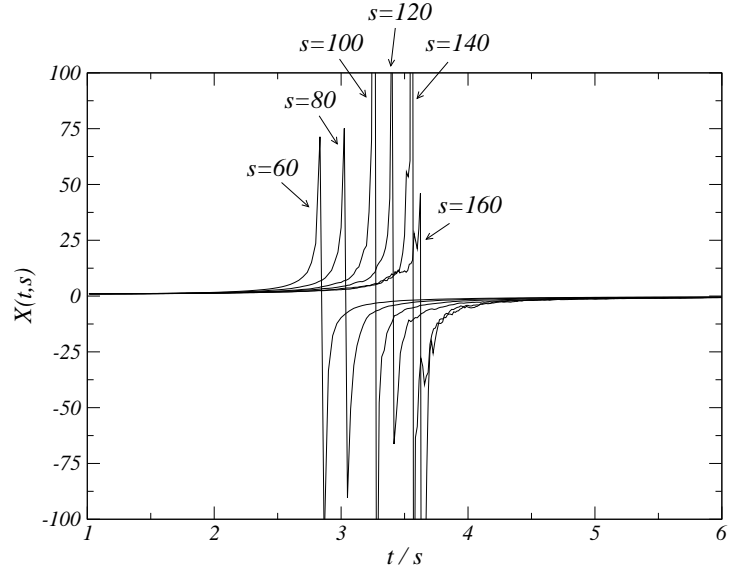}}
        \caption{Fluctuation-dissipation ratio $X(t,s)$ at $T=0.2T_{KT}$.
	The different curves correspond to different waiting times $s$.}
       \label{fig5}
\end{figure}
\end{center}

\begin{center}
\begin{figure}
        \centerline{\psfig{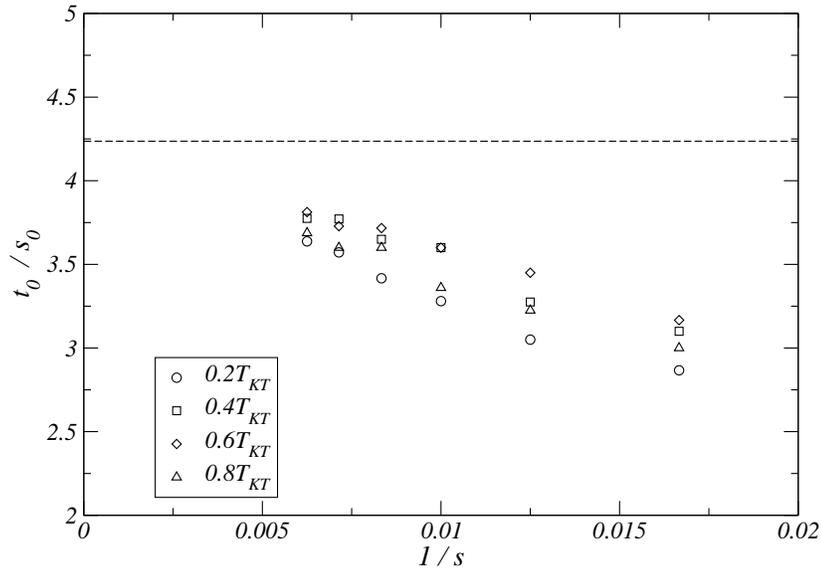}}
        \caption{Location $t_0/s_0$ of the divergence of the fluctuation-dissipation
	ratio $X(t,s)$ versus $1/s$. The different curves correspond to different
	temperature. The dashed line is the expected value in the spin-wave
	approximation.}
        \label{fig6}
\end{figure}
\end{center}

\section{Aging of angles}
\begin{center}
\begin{figure}
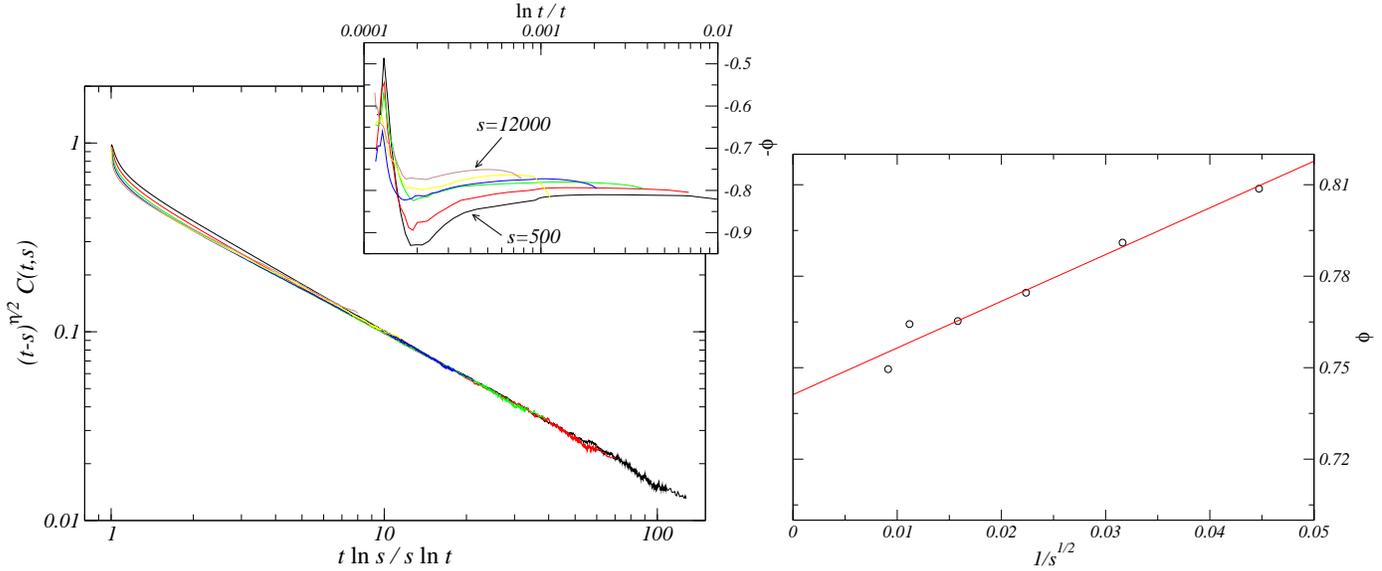

        \centerline{\psfig{figure=Corr-U1-HT-v2.eps,height=7.5cm}
	\psfig{figure=Exp-Eff-U1-HT.eps,height=5.5cm}}
        \caption{On the left, scaling function $(t-s)^{\eta/2}C(t,s)$ versus
	${t\ln s\over s\ln t}$ of the spin-spin correlation function for the FFXY
	prepared in a random state and quenched at the critical temperature. The
	different curves correspond to different waiting times ($s=500, 1000,
	2000, 4000, 8000,$ and $12000$). The inset corresponds to effective exponents
	calculated using a power-law fit in the window $[t_{\rm min};t_f]$ and plotted
	versus $\ln t_{\rm min}/t_{\rm min}$. On the right, the effective
	exponents on the plateau are plotted versus $1/\sqrt s$. The red line is
	a linear fit of the data.}
        \label{fig8}
\end{figure}
\end{center}

We now consider the following protocol: the system is prepared in a random state, i.e.
at infinite temperature, and then quenched at the Kosterlitz-Thouless transition temperature.
As we will show, the dynamics of the FFXY shows aging in a way similar to the XY model.
Above the Kosterlitz-Thouless transition temperature, the spin-wave approximation is not
expected to hold anymore but we can assume, like in the XY model, that the two-time
spin-spin correlation function $C(t,s)$ decay as
	\begin{equation}
	C(t,s)\sim (t-s)^{-\eta/2}f\left(\xi(t)\over\xi(s)\right)
	\sim (t-s)^{-\eta/2}f_{\cal C}\left({t\ln s\over s\ln t}\right)
	\label{eq10}
	\end{equation}
The prefactor $(t-s)^{-\eta/2}$ corresponds to the quasi-equilibrium regime while the
scaling function $f$ encodes the aging behavior of the correlation function.
According to the aging hypothesis, the relevant scaling variable is $\xi(t)/\xi(s)$ where
the characteristic length of domains is expected to grow as $\xi(t)\sim (t/\ln t)^{1/z}$,
i.e. algebraically (infinite relaxation time) with a logarithmic correction due to
the pinning of the domain walls by the free vortices present in the paramagnetic
phase~\cite{Bray00}. In the case of the XY model, the dynamical exponent is
$z=2$. To test eq. (\ref{eq10}), we have performed Monte Carlo
simulations for a lattice size $256\times 256$ with a final time $t_f=60,000$. The data
have been averaged over 5000 histories. The data are noisier than in the low-temperature
and it is not possible to obtain a more accurate estimate of $\eta$ from the quasi-equilibrium
regime. On figure \ref{fig8}, the scaling function $f_{\cal C}=(t-s)^{\eta/2}C(t,s)$ is
plotted versus ${t\ln s\over s\ln t}$. The different curves correspond to waiting times
$s=500, 1000, 2000, 4000, 8000,$ and $12000$. The collapse of the different curves has been
obtained using $a_c=0.1$. We are left with an algebraic decay $f_{\cal C}(x)\sim x^{-\phi}$
that confirms the scaling hypothesis (\ref{eq10}). A close inspection of the effective
exponents for the different waiting times reveals an evolution with $s$. On the
inset of fig \ref{fig8} the effective exponents calculated using a power-law fit
$f_{\cal C}(x)\sim x^{-\phi}$ in the window $[t_{\rm min};t_f]$ are plotted versus
$\ln t_{\rm min}/t_{\rm min}$. The different curves correspond to the different waiting
times $s$. Again, the large fluctuations for the leftmost points (larger values of $t$)
occur at the same value of $t$, independently of the waiting time $s$. We thus interpret
them as due to an under-sampling of the correlation causing systematic deviations of the
effective exponents when the number of points entering into the interpolation becomes small.
As a consequence, we have to consider the plateaus. An evolution with $s$ of the
effective exponents on these plateaus is clearly perceptible. The exponents are plotted
on figure \ref{fig8}. The extrapolation for $s\rightarrow +\infty$ leads to $\phi\simeq 0.74$
and thus, defining the autocorrelation exponent as~\cite{Huse89} $C(t,s)\sim t^{-\lambda/z}
\sim t^{-\eta/2-\phi}$, we obtain $\lambda/z\simeq 0.84$. This value is slightly larger
than the one obtained in the short-time regime from the initial-slip exponent $\theta$
($\lambda/z={d\over z}-\theta=0.808(3)$)~\cite{Luo98}. Note that short-time dynamics is
equivalent to our analysis with $s=0$~\cite{Oliveira} and is thus unable to give
a correct estimate if the effective exponent evolves strongly with the waiting time
$s$. Both our estimate of $\lambda/z$ and that obtained by Short-Time Dynamics
are incompatible with the autocorrelation exponent of the XY model:
$\lambda/z=0.625$~\cite{Luo98}, $0.738(4)$~\cite{Zheng03} or $0.738(5)$~\cite{Lei07}.

\begin{center}
\begin{figure}
        \centerline{\psfig{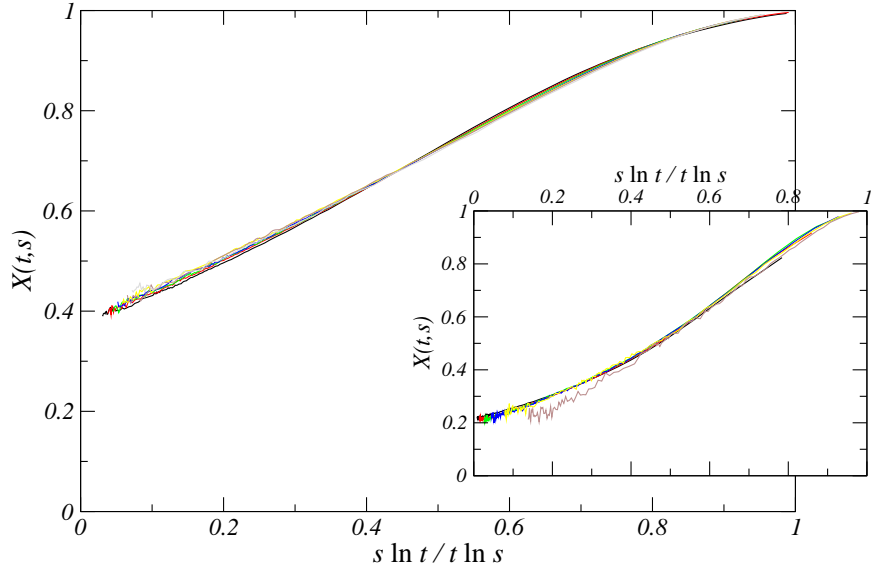}}
        \caption{Fluctuation-dissipation ratio associated to the angles.
	for the FFXY prepared in a random state and quenched at the critical temperature. 
	The different curves correspond to different waiting times ($60, 80, 100, 120, 140, 160,$
	and $180$). In the insert, the FDR is plotted for the XY model under the same
	protocol. The different curves correspond to $s=10, 20, 50, 100, 200,$ and $400$.}
        \label{fig9}
\end{figure}
\end{center}

Again the response is too noisy to investigate its asymptotic scaling behavior but,
using smaller lattice sizes and final times, it is still possible to improve significantly
the statistics and estimate accurately the fluctuation-dissipation ratio. We have used
a lattice size $192\times 192$ with time running up to $t_f=4,000$.
This allowed us to average the data over $500,000$ histories. On figure \ref{fig9}, the
FDR is plotted with respect to ${t\ln s\over s\ln t}$. A good collapse of the data
for different waiting times is observed. This collapse is not as good if logarithmic
corrections are omitted. The asymptotic value of the FDR is estimated to be $X_\infty=0.385(15)$.
In contradistinction to the algebraic decay of the spin-spin correlation function,
$X_\infty$ does not present any evolution with the waiting time $s$.
For comparison, we have performed a similar simulation for the pure XY model. The
asymptotic FDR is estimated to be $X_\infty=0.215(15)$ for the XY model, i.e.
very different from our value for the FFXY.

\def\square{\mathop{\vbox{\hrule height.4pt\hbox{\vrule width.5pt height3.5pt
\kern 3.5pt\vrule width.5pt}\hrule height.5pt}}\nolimits}

\section{Aging of the chirality}
We have also investigated the dynamics of the chirality, denoted $\chi$ in the following.
For each plaquette, the local chirality is defined as
	\begin{equation}
	\chi_{\square}={\rm sign}\ \Big[\sum_{(i,j)\in\square} J_{ij}\sin(\theta_i-\theta_j)\Big]
	\end{equation}
where the sum extends over the four bonds forming the plaquette.
Chirality is expected to behave as antiferromagnetically-coupled Ising spins
but a controversy exists regarding the belonging of the associated transition to
the Ising universality class. In the following, we will compare two-time chirality-chirality
correlation functions and the FDR associated to the response to a field coupled to
the chirality with those of the Ising model.

\begin{center}
\begin{figure}
        \centerline{\psfig{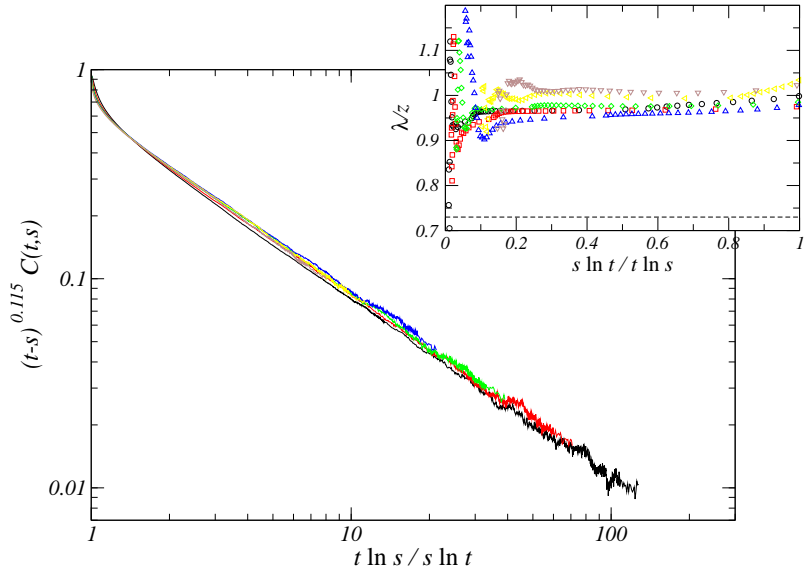}}
        \caption{Chirality-chirality correlation function $C(t,s)$ versus $t\ln s/s\ln t$
	for the FFXY prepared in a random state and then quenched at the inverse critical
	temperature $T_c=0.4545$. The different curves correspond to different waiting
	times ($500, 1000, 2000, 4000, 8000,$ and $12000$). In the insert, the effective
	exponent $\lambda/z$ computed by a power-law interpolation
	$(t-s)^{a_c}C(t,s)\sim (t/\ln t)^{-\lambda/z+a_c}$ between $t$ and
	$t_f$ is plotted with respect to $[t\ln s/s\ln t]^{-1}$. The dashed line
	corresponds to the exponent $\lambda/z$ for the pure model.}
        \label{fig10}
\end{figure}
\end{center}


The system is initially prepared in the paramagnetic phase and then quench at the $Z_2$
critical temperature. We have considered two different estimates of the critical
temperature: $T_c=0.4545(2)$~\cite{Luo98b,Luo98a} and $T_c=0.453243(2)$~\cite{Hasenbusch05}.
We have computed the two-time chirality-chirality correlation function
	\begin{equation}
	C(t,s)=\langle\chi(t)\chi(s)\rangle
	\end{equation}
for a lattice size $256\times 256$ up to the final time $t_f=200,000$. Data have
been averaged over 5000 histories. During a quench at the critical temperature,
autocorrelation functions are expected to behave as $C(t,s)\sim (t-s)^{-a_c} f(\xi(t)/\xi(s))$
where the first term corresponds to the quasi-equilibrium regime characterized
by a critical exponent $a_c=2\beta/\nu z_c$ and the second one is the aging part depending
only on the ratio $\xi(t)/\xi(s)$. In the FFXY, we assume that the existence of free
topological defects in the high-temperature induces logarithmic corrections in the growth
law. Thus the correlation length for the chirality-chirality correlations is expected to
grow as $\xi(t)\sim (t/\ln t)^{1/z_c}$, i.e. in the same way as for the spin-spin correlation
functions. As a consequence, we expect the scaling behavior
	\begin{equation}
	 C(t,s)\sim (t-s)^{-a_c}f_{\cal C}\left({t\ln s\over s\ln t}\right)
	\end{equation} 
This behavior of the chirality-chirality correlation function is observed numerically
only with the estimate $T_c=0.4545$ of the critical temperature. Figure~\ref{fig10}
shows the collapse of the scaling function $f_{\cal C}=(t-s)^{a_c}C(t,s)$
versus $t\ln s/s\ln t$ for different waiting times. This collapse is clearly destroyed
when $a_c$ is chosen smaller than $0.08$ or larger than $0.12$. This leads us to
$a_c=0.10(2)$, compatible with the Ising value $a_c=2\beta/\nu z_c\simeq 0.115$.
We note that removing the logarithmic corrections does not affect significantly the
quality of the collapse. For the second and most accurate estimate of the critical
temperature $T_c=0.453243(2)$, no collapse of the scaling function is obtained
for any positive value of $a_c$ (only a slightly negative value would bring the data
to collapse). This shows that the system is not critical in the quasi-equilibrium
regime but already in the low-temperature phase. In the following, we will consider
only the estimate $T_c=0.4545$ of the critical temperature.
\\

As shown on figure \ref{fig10}, the aging part $f_{\cal C}(x)$ of the chirality-chirality
correlation function decays algebraically. To recover the usual form $C(t,s)\sim s^{-a_c}
\big(t/s\big)^{-\lambda/z}$ in the asymptotic regime, the scaling function $f_{\cal C}(x)$
needs to decay as $x^{-\lambda/z+a_c}$. In the inset of figure \ref{fig10}
is plotted the effective exponent $\lambda/z$ computed by a power-law interpolation in
the interval $[t,t_f]$ where $t$ runs from $s$ to $t_f-2$. This effective exponent is
plotted with respect to ${s\ln t\over t\ln s}=x$ for the different waiting times $s$.
The exponent $\lambda/z$ is recovered in the limit $x\rightarrow 0$, i.e. at the left
side of the graph. Of course, in this region the number of points entering in the
interpolation is small and thus the accuracy of the effective exponent decreases.
Moreover, the data have been produced by a slow dynamics and are thus correlated. The
smooth variations at the left are really fluctuations and not any physical effect. From
this plot, we estimate $\lambda/z=0.98(5)$, i.e. incompatible with the Ising value
$\lambda/z\simeq 0.738$ (dashed line on the graph). Even though the curve of the
effective exponent for the largest waiting time is above the others, this cannot be
interpreted as an evolution of the exponents with $s$. Indeed, the sequence of
curves is not ordered according to the waiting times. Our conclusion
is remarkably different from that drawn for the Ising-XY model in the low-temperature phase
for which an estimate $\lambda\simeq 1.25$ compatible with that of the pure Ising
model was measured already for a relatively small final time $t_f=1000$~\cite{Kim96}.
For completeness, we should note that when we remove the logarithmic corrections,
the effective exponent is smaller, $\lambda/z\simeq 0.90(5)$ but still
incompatible with the Ising value. However, we believe that logarithmic corrections
are necessary as we will see in the following when considering the scaling of
the fluctuation-dissipation ratio.
\\

\begin{center}
\begin{figure}
        \centerline{\psfig{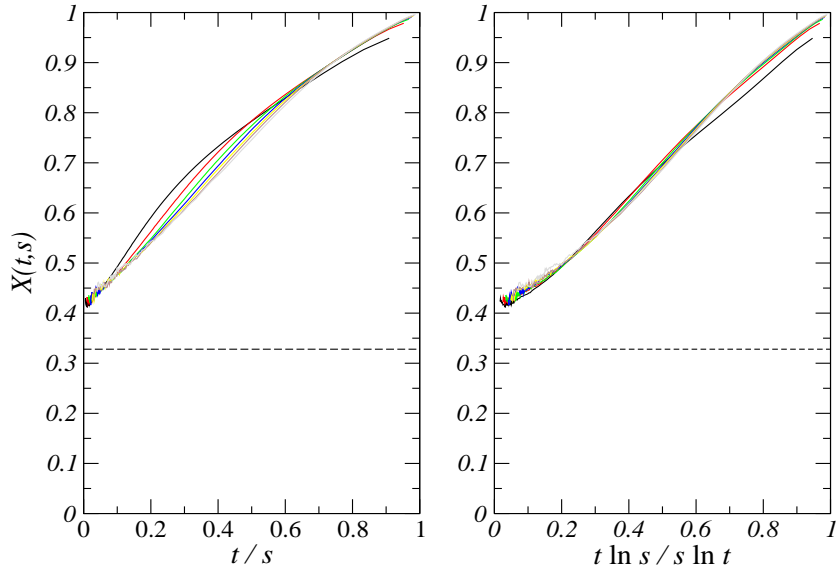}}
        \caption{Fluctuation-Dissipation Ratio $X(t,s)$ associated to the chirality
	versus $t/s$ (left) and $t\ln s/s\ln t$  for the FFXY prepared in a random state
	and quenched at the critical temperature. The different curves correspond to
	different waiting times ($10, 20, 30, 40, 50, 60,$ and $80$). The dashed line
	corresponds to the asymptotic value for the Ising model.}
        \label{fig11}
\end{figure}
\end{center}

The fluctuation-dissipation theorem is also violated at the $Z_2$ phase transition.
A Zeeman Hamiltonian
	\begin{equation}
	-\sum_{\square} h_{\square}\chi_{\square}
	\end{equation} 
that couples the chirality of each plaquette to an external field $h_{\square}$,
allows for the definition of a response function
	\begin{equation}
	R(t,s)=\left({\delta\langle\chi_{\square}(t)\rangle
	\over\delta h_{\square}(s)}\right)_{h_{\square}\rightarrow 0}
	\end{equation} 
This quantity is evaluated numerically~\cite{Chatelain03}. We can then compute
the fluctuation-dissipation ratio $X(t,s)=k_BT R(t,s)/\partial_s C(t,s)$.
We used a lattice size $192\times 192$ with a final time $t_f=2000$.
Data have been averaged over $10^6$ histories. The FDR is plotted on figure
\ref{fig11} with respect to $s/t$ (left) and $s\ln t/t\ln s$ (right).
A better collapse is obtained with logarithmic corrections, showing that
the dynamics of the domain walls between different antiferromagnetic orderings
of the chirality is sensible to the vortices. The asymptotic value $X_\infty$
is estimated to $X_\infty=0.405(5)$ (a compatible value would be obtained
if logarithmic corrections would not be taken into account). This value is
significantly different from the Ising value $X_\infty\simeq 0.328$~\cite{Chatelain04}.
Again, no evolution of $X_\infty$ with the waiting time $s$ is observed.

\section*{Conclusions}
We have given numerical evidences that the thermal excitations in the low-temperature
phase are spin-waves whose dynamics is the same as in the XY model. The dynamical
exponent is thus $z=2$ in the whole critical phase. The exponent $\eta$
that can be estimated from the spin-spin two-time autocorrelation function is
compatible with the prediction of the spin-wave approximation and with estimates from
static Monte Carlo simulations close to $T_{KT}$. The fluctuation-dissipation ratio
displays divergences, like in the XY model, at some locations that tend toward the value
predicted by the scaling theory. When initially prepared in the paramagnetic phase and
then quenched at $T_{KT}$ (resp. $T_c$), spins (resp. chiralities) age in the same
way as an unfrustrated ferromagnet. Spin-spin autocorrelation function decays slightly faster
($\lambda/z\simeq 0.84$) than in the XY model. The FDR takes asymptotically
a very different value too. These results are consistent with the fact that our
estimate of $\eta=0.27(2)$ is greater than the XY value, even though at the boundary
of the error bar. The chirality-chirality autocorrelation function decays algebraically
at the critical point with an exponent $\lambda/z\simeq 0.98(5)$ incompatible with
that of the Ising model. The evolution of the effective exponent with the waiting time
was taken into account. The FDR is also fully incompatible with the Ising value.
No evolution of these values with the waiting time is visible, in contradistinction
to static Monte Carlo simulations for which a cross-over is clearly observed when
increasing the lattice size~\cite{Noh02,Hasenbusch05}.
The coupling between angles and chiralities is probably still too strong in the
range of parameters used in our simulations. This coupling is manifest in the fact that
the FDR is affected by logarithmic corrections that may be interpreted as an influence
of the topological defects causing the Kosterlitz-Thouless transition. Moreover, the fact
that our estimate of $X_\infty$ for the chirality is relatively close to the value
measured at the Kosterlitz-Thouless temperature for the angles may also be the result
of a strong coupling. To reach the cross-over regime and eventually enter the true critical
regime, the correlation length should be allowed to reach much larger values. This
requires of course larger lattice sizes but also larger times since $\xi\sim ({t/\ln t})^{1/z}$.
The computational effort scales with the lattice size as $L^{2+z}$. It is therefore
identical to static Monte simulations since the number of Monte Carlo steps needs to
be increased when the lattice size increases and only local updates are possible for the
FFXY. However, it seems that the amplitude is playing against us!

\ack
The institute Jean Lamour is Unit\'e Mixte de Recherche CNRS number 7198.

\end{document}